\documentclass[article,floatfix,showpacs,onecolumn]{revtex4}
\usepackage{amsmath}
\usepackage{graphicx}

\begin{document}
\title{Coarse-graining diblock copolymer solutions: a macromolecular version of the Widom-Rowlinson model}

\author{C.I. Addison$\dagger$, J.P. Hansen*$\dagger$, V. Krakoviack$\ddagger$, A. A. Louis$\dagger$} \affiliation{$\dagger$ Dept. of Chemistry, University of
Cambridge, Lensfield Road, CB2 1EW, Cambridge, UK \\ \\ $\ddagger$Laboratoire de Chimie,
Ecole Normale Sup\'{e}rieure de Lyon, 69364 Lyon Cedex 07, France}

 \begin{abstract}
 We propose a systematic coarse-grained representation of block
 copolymers, whereby each block is reduced to a single ``soft blob''
 and effective intra- as well as intermolecular interactions act
 between centres of mass of the blocks. The coarse-graining approach
 is applied to simple athermal lattice models of symmetric AB diblock
 copolymers, in particular to a Widom-Rowlinson-like model where
 blocks of the same species behave as ideal polymers ({\it i.e.}\
 freely interpenetrate), while blocks of opposite species are mutually
 avoiding walks.  This incompatibility drives microphase separation
 for copolymer solutions in the semi-dilute regime.  An appropriate,
 consistent inversion procedure is used to extract effective inter-
 and intramolecular potentials from Monte Carlo results for the pair
 distribution functions of the block centres of mass in the infinite
 dilution limit.

 \end{abstract}
\pacs{61.25.Hq,61.20.Gy,05.20Jj}

 \maketitle

 \section[Intro]{Introduction}\label{intro}
Recent years have witnessed considerable theoretical interest in
statistical mechanics of multicomponent systems involving multiple
length and time scales.  To be tractable, statistical descriptions of
such multi-scale systems must resort to controlled coarse-graining
methods. One widely used coarse-graining strategy for complex fluids
is to determine effective interactions between large (``dressed'')
particles ({\it e.g.}\ macromolecules or colloidal particles) by
systematically tracing out the degrees of freedom of small particles
({\it e.g.}\ counter-ions or solvent molecules
\cite{likos}\cite{louis}\cite{hans1}). A particularly successful
application has been to dilute and semi-dilute polymer solutions:
effective interactions between polymer centres-of-mass (CM) are
determined by taking appropriate averages over monomer conformations
for fixed distances {\it r} between the $CM$s of interacting
linear\cite{hall}\cite{bolh}\cite{krak}\cite{Guen} or star
polymers{\cite{likos}\cite{jusu}.  In the case of linear polymers, to
which the subsequent discussion will be restricted, the resulting
$CM-CM$ pair potential for polymers in good solvent, modelled by self
avoiding walks (SAW) on a cubic lattice, is very soft, repulsive and
finite at all separations {\it r}, with a range of the order of the
radius of gyration $R{_g}$ of the polymer coils.  The general shape is
reasonably well represented by a Gaussian of amplitude on the order of
$k{_B}T$, reflecting the essentially entropic nature of the effective
interaction.  While the overall form of the effective pair potential
is not very sensitive to polymer concentration under good solvent or
high temperature conditions, it turns out to be very sensitive to
concentration and temperature when the latter is lowered toward the
$\theta$ solvent regime\cite{krak}\cite{pelis}. \\ Replacing a full
monomer level representation of polymer coils by a ``soft particle''
description based solely on effective pair potentials between $CM$s
represents a considerable reduction in the number of degrees of
freedom, and hence leads to a concomitant reduction in computational
effort in simulations of large-scale phenomena involving many
polymers.  Moreover, use of spherically symmetric effective pair
potentials allows a direct exploitation of the theoretical arsenal
developed over the years for the study of the bulk\cite{hans2} and
interfacial\cite{widom1} properties of simple fluids.  Last, but not
least, inspection of the effective pair potentials leads to new
insights into the phase behaviour\cite{louis} and stability\cite{krak}
of macromolecular solutions.\\ In this paper we extend the above
coarse-graining strategy to the case of solutions of block copolymers
and more specifically to symmetric diblock copolymers made up of two
strands A and B of equal length.  AB copolymers will be reduced to
``soft diatomics'', with inter- and intra-molecular interactions
between sites associated with the $CM$s of A and B strands.  The nature
of A-A, A-B and B-B effective pair interactions depends on solvent
conditions.  Most of the results presented in this paper will be based
on a simplified model inspired by the celebrated Widom-Rowlinson (WR)
model for fluid/fluid phase separation\cite{widom2}.  In the
macromolecular extension of the WR model, A and B strands are mutually
avoiding ({\it i.e.}\ monomers of opposite species cannot overlap)
while A and B strands will separately behave like ideal (Gaussian)
chains, {\it i.e.}\ they will freely interpenetrate same species
strands on different copolymers. \\

\section{Model and coarse-graining strategy}
We consider lattice models of polymers, where monomers occupy the
sites of a periodically repeated simple cubic lattice of size
$\Lambda^3$; the bond vectors connecting successive monomers point
along the {\it x}, {\it y} or {\it z} directions only.  Each polymer
comprises ${\cal M}$ monomers and hence $ L={\cal M}-1$ bonds (or
segments), such that the length of the polymer is $L{\it b}$, if {\it
b} is the segment length. A diblock copolymer AB is made up of two
strands comprising $M_A$ and $M_B$ monomers respectively, with the
$M{_A}${th} monomer of strand A connected to the first monomer of
strand B.  The total length is $(L_A + L_B + 1)b=(M_A + M_B -1)b$.
Restriction will be made here to symmetric copolymers, such that
$M_A=M_B=M $ and $L_A=L_B=L$.  We will consider N such copolymers on
an $\Lambda^3$ lattice, such that the A and B monomer concentrations
are $c_A=c_B=MN/\Lambda^3$ and the overall monomer concentration is
$c=c_A + c_B$.  Individual copolymer chains are characterised by three
radii of gyration $R_{gA}$, $R_{gB}$ and $R_{g}$, defined as usual by
\begin{eqnarray}
\label{eq1}
R_{gA}^2= \frac{1}{M_A}\sum^{M_A}_{i=1} <(\overrightarrow r_i^A - \overrightarrow R_A)^2>
\end{eqnarray}
where $\overrightarrow r_i^A$ is the position of the $i^{th}$ monomer
of species A on the lattice, while $\overrightarrow
R_A=\sum^{M_A}_{i=1} \overrightarrow r_i^A/M_A$ is the position of the
centre-of-mass ($CM$) of strand A. The
statistical average $<.>$ is taken over properly weighed chain
conformations.  Similar definitions hold for $R_{gB}$ and for the
overall radius of gyration $R_g$; the overall $CM$ is $\overrightarrow
R=(\overrightarrow R_A + \overrightarrow R_B)/2$.  Another important
length is the root mean square distance between the $CM$s of the two
strands of a copolymer:
\begin{eqnarray}
\label{eq2}
R_{AB}=[<|\overrightarrow R_A -\overrightarrow R_B|^2>]^{1/2}
\end{eqnarray}
An elementary calculation shows that:
\begin{eqnarray}
\label{eq3}
R_{g}^2=\frac{1}{2}R_{gA}^2 +\frac{1}{2}R_{gB}^2 + \frac{1}{4}R_{AB}^2.
\end{eqnarray}
The A-A, A-B and B-B interactions between non-adjacent monomers of the
same or opposite species determine the structure and phase behaviour
of individual copolymers and of many-copolymer systems.  A number of
models have been examined in the polymer literature: the more
interesting are those which account for some degree of incompatibility
between A and B monomers, which leads to microphase
separation\cite{bates}\cite{binder}.  In this paper we focus on the
simplest, athermal models combining ideal (or random walk, with
backward overlapping steps allowed) strands (I) and excluded volume (self or
mutually avoiding walks) strands (S).  The eight possible combinations
of monomer interactions will be labelled by three indices JKL, with
each index either I or S.  The first index refers to the interaction
between A monomers (on the same or different polymers), the second
index refers to the A-B cross interaction, while the last index refers
to the B-B interaction.  Of the eight possible combinations, two (III
and SSS) reduce to homopolymers of molecular weight $2M$, while IIS
and SII as well as ISS and SSI are degenerate for the symmetric
($M_A=M_B$) copolymers considered here, leaving four distinct,
non-trivial copolymer models, namely IIS, ISI, ISS and SIS.  Among
these, the ISI model may be regarded as the macromolecular equivalent
of the WR model\cite{widom2}.  The macroscopic phase separation
occurring in the latter is prevented here because A and B strands are
tethered, but microphase separation into a lamellar structure may be
expected.  Most of the subsequent discussion is hence devoted to the
ISI model. \\ Physically, self avoiding walks and random walks (ideal
polymers) reflect solvent conditions.  The former model corresponds to
good solvent conditions associated with swollen chains, while random
walk polymers are generally considered as approximating
$\theta$-solvent conditions, at least for low polymer
concentrations\cite{add1}.  Thus in the ISI model, A and B strands
behave individually as if they were in a $\theta$-solvent, where
monomer excluded volume effects are compensated by nearest-neighbour,
solvent-induced attraction, while A-B interactions are purely of
excluded volume nature (mutually avoiding walks).  A more general
model than those considered here contains both excluded volume
interactions (single occupancy constraint) and nearest neighbour
interactions $\epsilon_{\alpha\beta} (\alpha,\beta= A$ or $B$), thus
introducing an explicit temperature dependence.  This standard model
has been extensively studied both in the low concentration(single
copolymer) limit\cite{olaj}, and in the melt\cite{binder}.  The
present investigation covers special cases of the generic model for
finite copolymer concentrations, in the dilute and semi-dilute
regimes.  The key characteristic of these models, and in particular of
the ISI model, is their extreme non-additivity, as in the WR model. \\
The range of concentrations which will be considered is best
characterised by the ratio of the copolymer density, $\rho=N/(\Lambda
b)^3$ over the overlap density $\rho*=3/(4\pi R{_g}{^3})$, where $R_g$
is the radius of gyration at zero density.  Dilute solutions correspond to
$\rho/\rho*\ll1$, while the semi-dilute regime is for $\rho/\rho*\geq
1$.  \\ The coarse-graining strategy which we shall pursue generalises
that developed earlier for linear
homopolymers\cite{hall}\cite{bolh}\cite{krak}.  An AB block copolymer
will be represented by two spherical ``blobs'' centred on the $CM$s
${\bf R}_A$ and ${\bf R}_B$, and tethered by an anharmonic entropic
spring deriving from an intramolecular potential $\phi_{AB}(r)$, where
$r=|{\bf R}_A-{\bf R}_B|$.  The A and B blobs on different copolymers
will interact via effective pair potentials $v_{AA}(r), v_{AB}(r)$ and
$v_{BB}(r)$ acting between the $CM$s. As in the earlier work on
homopolymers, these 4 effective potentials are determined by an
inversion procedure based on the $CM-CM$ pair distribution functions
$s_{AB}(r)$ (intramolecular) and $g_{AA}(r), g_{AB}(r)$ and
$g_{BB}(r)$ (intermolecular).  The latter are determined by Monte
Carlo (MC) simulations of the initial, monomer-level model. \\ In this
paper these pair distribution functions and resulting effective pair
potentials will be determined only in the low concentration limit, by
simulating copolymer pairs.  While the inversion procedure is trivial
in that limit in the case of homopolymers(represented by single
blobs), it turns out to be much more involved in the copolymer case,
since four effective blobs are involved, as will be discussed in more
detail in section IV.

\section{Monte Carlo Results} 
With the above coarse-graining objective in mind, we have carried out
extensive MC simulations of symmetric AB diblock copolymers on a cubic
lattice.  Most simulations were for the ISI model, but some
preliminary results have also been obtained for the IIS, ISS and SIS
models; test runs were carried out for the SSS model to compare with
previous results of the equivalent SAW homopolymers. \\ MC runs were
carried out on periodic lattices of size at least $\Lambda=100$.
The numbers of copolymers were adjusted to achieve any given polymer
density $\rho/\rho*$; thus $N$ was taken to be 1 to determine infinite
dilution, single polymer properties, $N=2$ if effective pair
interactions were to be determined in that limit, while for
$\rho/\rho* \geq 1$, $N$ varied between a few hundred and over a
thousand.  The number of monomers $2M$ in each strand of a copolymer was
varied between 60 and 1000.  Configuration space was sampled by using
pivot and translation moves\cite{lal}\cite{madras}, polymers were
subjected to between $5\times10^5$ and $10^8$ MC moves, depending on the
total number of polymers in the system. \\
\begin{table}[!htp]
\begin{center}
\begin{tabular}{l||c|c|c|c|c|}	
\multicolumn{6}{l}{\rule[-2mm]{0mm}{6mm}2M=320}\\
\hline 
{$Type$}&    {$R_g$}    & {$R_{gA}$}&{$R_{gB}$} & $R_{AB}$ & 
\\ \hline
ISI     &       8.19  & 5.27 &  5.28 & 14.11 &
\\
IIS     &       9.99  & 5.16 &  8.53 & 12.47 & 
\\
ISS     &      10.85  & 5.27 &  8.69 & 16.08 & 
\\ 
SIS     &      12.13  & 8.55 &  8.50 & 17.11 & 
\\
SSS     &      12.96  & 8.70 &  8.64 & 19.14 & 
\\
\hline
\multicolumn{6}{l} {\rule[-2mm]{0mm}{6mm}2M=500} \\
\hline
{$Type$}&    {$R_g$}    & {$R_{gA}$}&{$R_{gB}$} & $R_{AB}$ & $B_2$
\\ \hline
ISI	&      10.228  & 6.588 &  6.586  & 18.23  & 4.84
\\
IIS     &      12.902  & 6.455 &  11.146 & 15.60  & 3.76
\\
ISS     &      13.872  & 6.560 &  11.279 & 20.66  & 5.71
\\ 
SIS     &      15.798  & 11.148 & 11.146 & 22.34  & 3.87
\\
SSS     &      16.831  & 11.273 & 11.271 & 24.96  & 5.70
\\\hline
\multicolumn{6}{l} {\rule[-2mm]{0mm}{6mm}2M=2000} \\
\hline
{$Type$}&    {$R_g$}    & {$R_{gA}$}&{$R_{gB}$} & $R_{AB}$ &
\\ \hline
ISI	&      20.47  & 13.20 & 13.18 &  40.32 &
\\
IIS     &      28.94  & 13.13 & 25.54 &  31.28 &
\\
ISS     &      30.33  & 13.13 & 25.59 &  45.23 &
\\ 
SIS     &      35.95  & 25.38 & 25.36 &  50.44 &
\\
SSS     &      38.39  & 26.06 & 26.04 &  56.23 &
\\ \hline
\end{tabular}
\end{center}
\caption{Radii of gyration $R_g$, mean intramolecular distance $R_{AB}$ and second virial coefficient $B_2$ in units of $R_g$ (2M=500 only) for ideal/SAW systems.}
\label{tab:rgtable3}
\end{table}
Results for the 3 radii of
gyration and the mean $CM-CM$ distance $R_{AB}$ are collected in Table
1, for the four copolymer models as well as for the SSS model, which
serves as a check against the well documented SAW homopolymer model,
for which $R_g\simeq0.44L^{\nu}$ with $\nu\simeq0.587$ in the scaling
limit\cite{li}. The trends in the various quantities in Table 1 for
the various models agree qualitatively with expectation.  The internal
CM-CM pair distribution function $s_{AB}$ and the resulting ``entropic
spring'' potential {\it i.e.}\ the effective bonding potential between
the $CM$s of the A and B strands of a single copolymer:
\begin{eqnarray}
\label{eq4}
\phi_{AB}(r) = -k_BT \ln s_{AB}(r) 
\end{eqnarray}
are plotted in figure \ref{fig1}.
\begin{figure}[!htp]
\begin{center}
\includegraphics[width=12cm,angle=0]{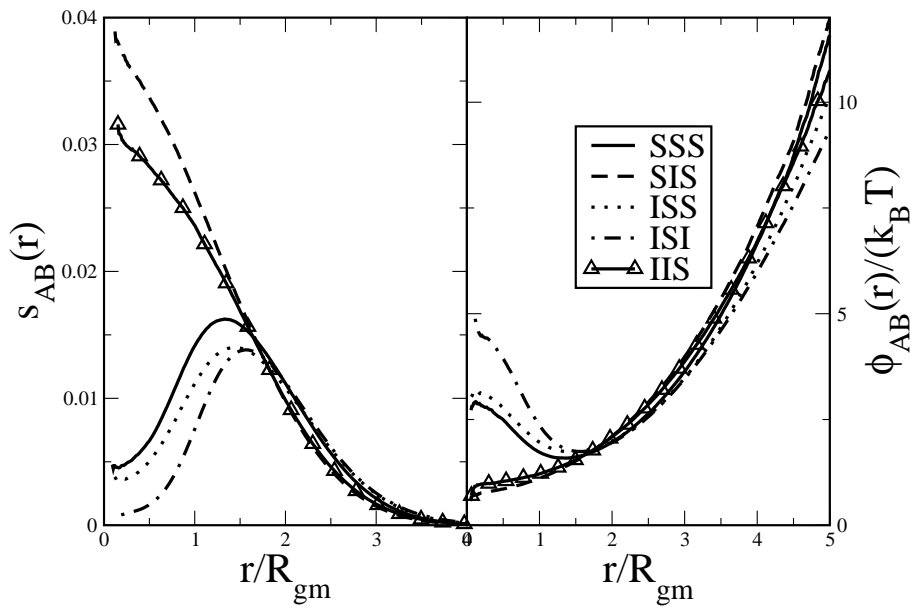}
\caption{{\bf left:} Zero density limit of the intramolecular $CM$-pair
distribution function $s_{AB}(r)$ for different models, versus
$r/R_{gm}$, where $R_{gm}=(R_{gA}+R_{gB})/2$ is the mean radius of
gyration of the A and B strands. {\bf right:} The corresponding
intramolecular pair potentials $\phi_{AB}(r)$, in units of $k_BT$,
defined in equation \protect\ref{eq4}.}
\label{fig1}
\end{center}
\end{figure}
While the distribution functions
$s_{AB}(r)$ for the IIS and SIS models peak at the origin, as one
might expect since both strands can freely interpenetrate in both
cases, the $s_{AB}(r)$ have peaks at $r> R_{gm}$, where
$R_{gm}=(R_{gA}+R_{gB})/2$, in the ISI, ISS and SSS models, because of
the single occupancy constraint on monomers of opposite species in all
three cases. These three copolymer models behave hence like
macromolecular diatomics, {\it i.e.}\ they are, on average, elongated,
as confirmed by inspection of the eigenvalues of their tensor of
inertia (not shown here).  The corresponding intramolecular potentials
$\phi_{AB}$ exhibit a maximum at full overlap ($r=0$) and a minimum
for $r\simeq1.5R_{gm}$.  Beyond that minimum, all potentials increase
rapidly, illustrating the ``entropic spring'' character of the
effective interaction between the $CM$s of the two strands. Because the
two strands are tethered, $\phi_{AB}$ must diverge when $r\rightarrow
L\times b$, corresponding to the highly improbably fully stretched
conformation of the AB copolymers. Note that relative to the $r=0$
value, the potential minimum is considerably deeper for the ISI model,
compared with the ISS and SSS models.  In other words elongated
conformations have higher probability for the ISI model.  This is
because the two ideal strands in the latter model are more compact
than self avoiding strands, as reflected in their shorter radii of
gyration.  The potential barrier to interpenetration of mutually
avoiding strands is thus larger when the strands are individually
ideal, since the probability of the forbidden overlap of monomers of
opposite species is larger.  The results discussed so far are for
isolated copolymers. We henceforth restrict all further considerations
of many polymer systems to the ISI model. \\
\begin{figure}[!htp]
\begin{center}
\includegraphics[width=12cm,angle=0]{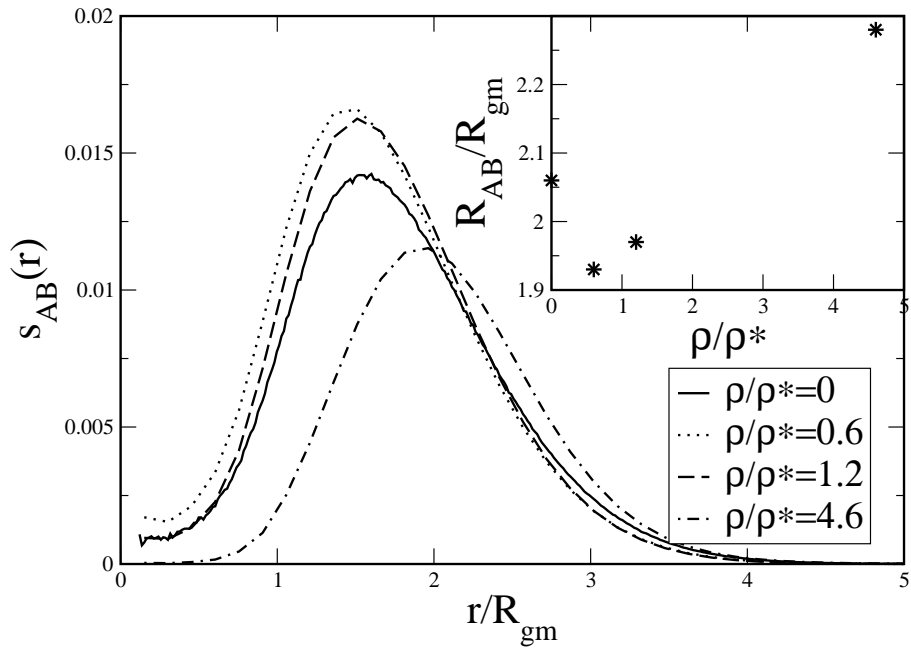}
\caption{Intramolecular $CM$ pair distribution function $s_{AB}(r)$ of the ISI model as a function of $r/R_{gA}$ at four densities $\rho/\rho*$. The inset shows $R_{AB}$ (defined in equation \protect\ref{eq5}) versus $\rho/\rho*$.}
\label{fig2}
\end{center}
\end{figure}
We first investigate the influence of polymer concentration on the
intramolecular $CM$ distribution function $s_{AB}(r)$.  Figure
\ref{fig2} shows the MC data for the ISI model at four different
concentrations, as well as the corresponding $R_{AB}$, which is
related to the properly normalised $s_{AB}(r)$ by:
\begin{eqnarray}
\label{eq5}
  R_{AB}^2 = 4\pi\int_{0}^{Lb} s_{AB}(r) r^4 dr.
\end{eqnarray}
 $s_{AB}(r)$ is seen to be rather insensitive to concentration in the
dilute regime as one might intuitively expect, $R_{AB}$ contracts
slightly as $\rho/\rho*$ increases before expanding at higher
densities where microphase separation sets in. In terms of the
coarse-grained diatomic analogy, this means that the two-blob
``molecule'' hardly contracts upon compression, and that this
coarse-grained entity remains a valid concept in the semi-dilute
regime.  Note however that the ``entropic spring'' potential
$\phi_{AB}$ is no longer given by eq.\ref{eq4} at finite
concentration, because the problem ceases to be a purely two-body
one. A more complex inversion procedure is required, to which we
return in section IV. \\ In order to gain more insight into the
physical significance of the ISI model, we have determined the osmotic
equation of state of the model in the dilute and semi-dilute regimes.
The equation of state can be efficiently and accurately determined in
a single MC run, by subjecting the copolymers to a gravitational field
until sedimentation equilibrium is reached, determining the resulting
inhomogeneous monomer or CM concentration profile $\rho(z)$ (where $z$
is the altitude) and extracting the osmotic pressure $P$ from the
measured profile by invoking hydrostatic equilibrium\cite{biben}\cite{add2}
\begin{eqnarray}
\label{eq6}
\frac{dP(z)}{dz}=-mg\rho(z)
\end{eqnarray}
\begin{figure}[!htp]
\begin{center}
\includegraphics[width=9cm,angle=0]{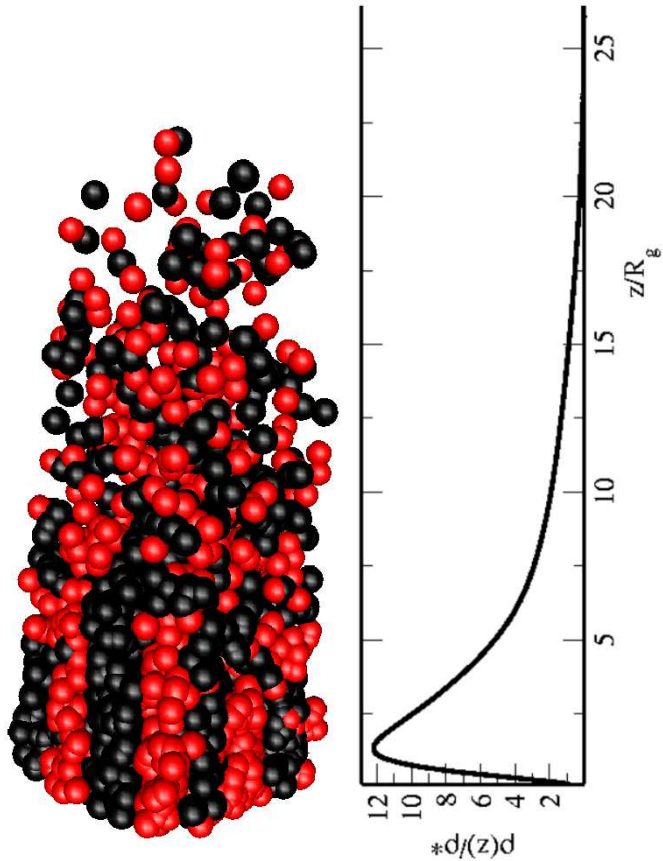}
\caption{{\bf Left:} typical configuration of N=1600 ISI copolymers under
gravity ($\zeta=k_BT/mg=1.66R_g$). The copolymers are confined to a
vertical box, open-ended in the $z$-direction($z>0$), and periodically
repeated in the horizontal $x$ and $y$ directions, the horizontal base is
100$\times$100 lattice units.  A and B strands are
pictured as black and grey, spheres of radius $R=R_{gA}$.  Microphase
separation into a striped pattern in clearly visible at lower
altitudes. {\bf Right:} monomer density profile $\rho(z)$ for the
system, as shown left.}
\label{fig3}
\end{center} 
\end{figure}
\begin{figure}[!htp]
\begin{center}
\includegraphics[width=8cm,angle=270]{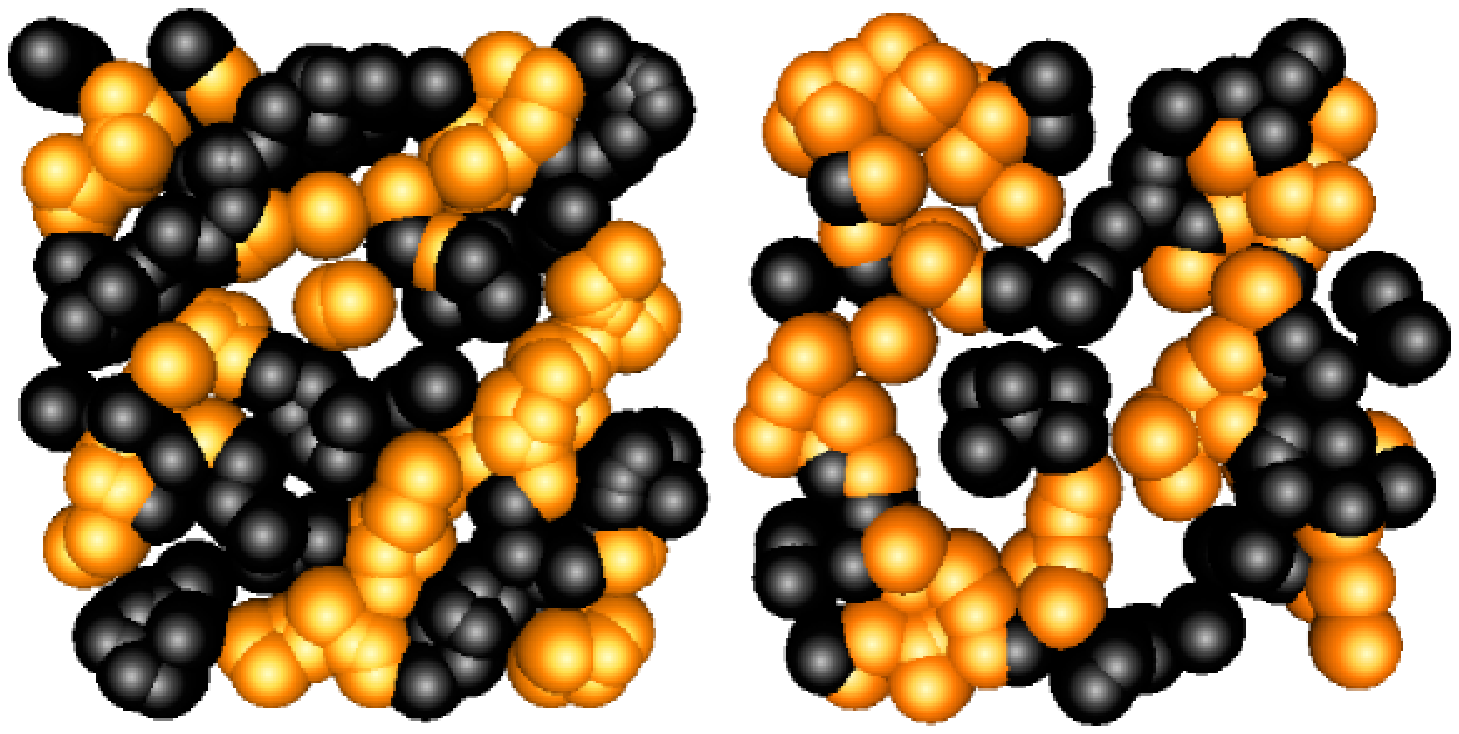}
\caption{Two horizontal cuts of width $\Delta z=R_{gA}$ of the
configuration in figure \protect\ref{fig3}, at altitudes
$z=5.5R_g$(left) and $z=9.5R_g$(right).  A and B strands are again
pictured as black and grey spheres of radius $R_g$. The lamellar structure is starting to break up at $z=9.5R_g$.}
\label{fig4}
\end{center} 
\end{figure}
where m is a fictitious buoyant mass and $g$ the acceleration due to gravity;
these parameters may be adjusted in the simulation, such that the
resulting sedimentation length $\zeta=k_BT/mg$ is larger than the
polymer radius of gyration $R_g$, thus ensuring the validity of the
macroscopic equation \ref{eq6} on the mesoscopic scale set by
$R_g$. A snapshot of a typical configuration in the sedimentation
column is shown in figure \ref{fig3}, while the concentration profile
$\rho(z)$ averaged over millions of such configurations generated in
the MC run is plotted adjacent in the figure.  The configuration in
figure \ref{fig3} clearly hints at the existence of microphase separation of
the ISI model for copolymer concentrations $\rho/\rho* \geq2$.
Horizontal cuts through the sedimentation column, shown in figure
\ref{fig4}, exhibit alternating stripes of A and B strands,
characteristic of the lamellar phase observed in many symmetric block
copolymer melts\cite{bates}\cite{binder}.  However, while the control
parameter in melts is the temperature, the control parameter for the
athermal ISI model is the polymer concentration. \\
\begin{figure}[!htp]
\begin{center}
\includegraphics[width=12cm,angle=0]{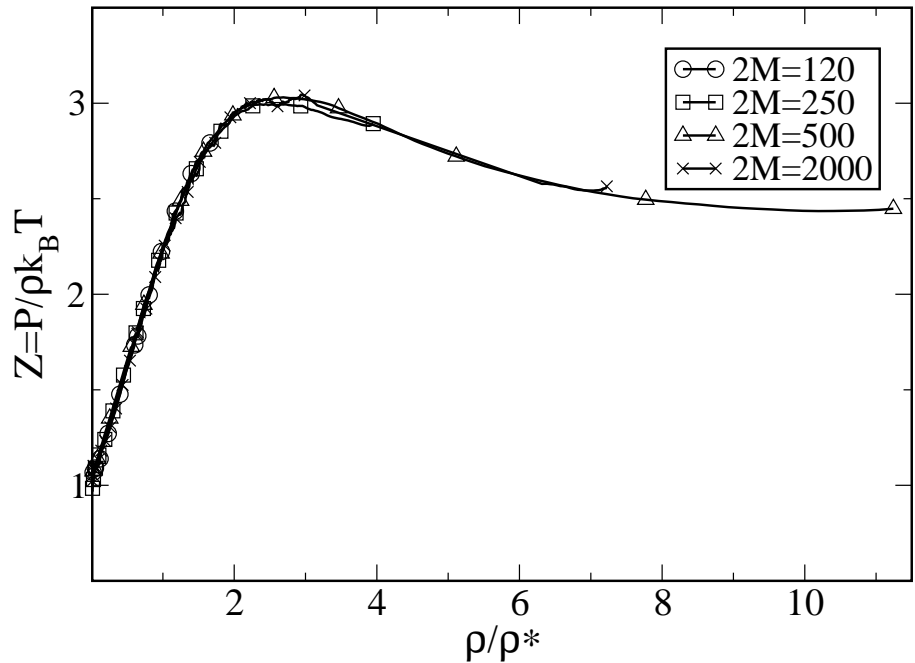}
\caption{Osmotic equation of state $Z=P/(\rho k_BT)$ of the ISI model as a function of $\rho/\rho*$ as extracted from equation \protect\ref{eq6}.  The results are for copolymers of different lengths 2M=120, 250, 500 and 2000}
\label{fig5}
\end{center} 
\end{figure}
Further evidence for the microphase separation comes from the osmotic
equation of state, extracted from the concentration profile as
explained above.  The dimensionless equation of state $Z=P/\rho k_BT$,
computed for copolymers with total number of monomers $2M$=120, 250,
500, and 2000 are plotted versus $\rho/\rho*$ in figure \ref{fig5}; no
significant size dependence of $Z$ is seen.  $Z$ increases from its
infinite dilution limit 1 up to $\rho/\rho* \approx 2.5$, where it
reaches a maximum, before decreasing slowly to what appears to be an
asymptotic value for $\rho/\rho* >10$.  With increasing density the system 
moves from a disordered region to one where the lamellar order increases, 
and there are fewer interactions 
between A and B strands.  The excess pressure(compared to ideal polymers) is 
decreases in this region, as a consequence of the lamellar ordering minimises 
the number of A-B overlaps.  The initial slope of $Z(\rho)$ agrees well with 
the second virial coefficient reported in table 1, while at high 
concentrations, the osmotic pressure increases 
sub-linearly with $\rho/\rho*$.  In future work the anisotropic pair structure 
will be calculated to characterise the onset of microphase separation in more detail.

\section{Solving the inverse problem}
 We now return to the main objective of the present paper, namely to
 determine the effective pair potentials $v_{AA}(r)$, $v_{AB}(r)$ and
 $v_{BB}(r)$ between the $CM$s of A and B strands on different
 copolymers.  In the case of the symmetric ISI model,
 $v_{AA}(r)=v_{BB}(r)$.  From earlier work on homopolymers\cite{bolh},
 which have coarse-grained interaction characterised by a single
 effective pair potential, we expect the effective pair potentials to
 be state dependent {\it i.e.}\ to depend on polymer concentration.
 The subsequent discussion will be restricted to the limit of
 vanishing concentration, {\it i.e.}\ to an isolated pair of
 copolymers.  To this end the intramolecular pair distribution
 functions $g_{AA}(r)=g_{BB}(r)$ and $g_{AB}(r)$ have been computed by
 MC simulations of a pair of AB copolymers.  The procedure is similar
 to that used previously for homopolymers\cite{bolh}\cite{krak},
 whereby conformations of the two isolated polymer coils are
 generated, and the $CM$s of the two coils are moved toward each other,
 while histograms are collected of the allowed configurations as a
 function of the $CM-CM$ distance.  In the case of diblock copolymers,
 care must be taken to correctly generate the partial
 $g_{\alpha\beta}(r)$.  The simulation must be adjusted to move
 polymers such that the relevant $CM_{\alpha}-CM_{\beta}$ distance is
 changed.  The MC results for the three low density pair distribution
 functions of the ISI model are shown in figure \ref{fig6} together
 with the pair distribution functions of the overall $CM$s of the
 copolymers, $g_{CC}(r)$. 
\begin{figure}[!htp]
\begin{center}
\includegraphics[width=12cm,angle=0]{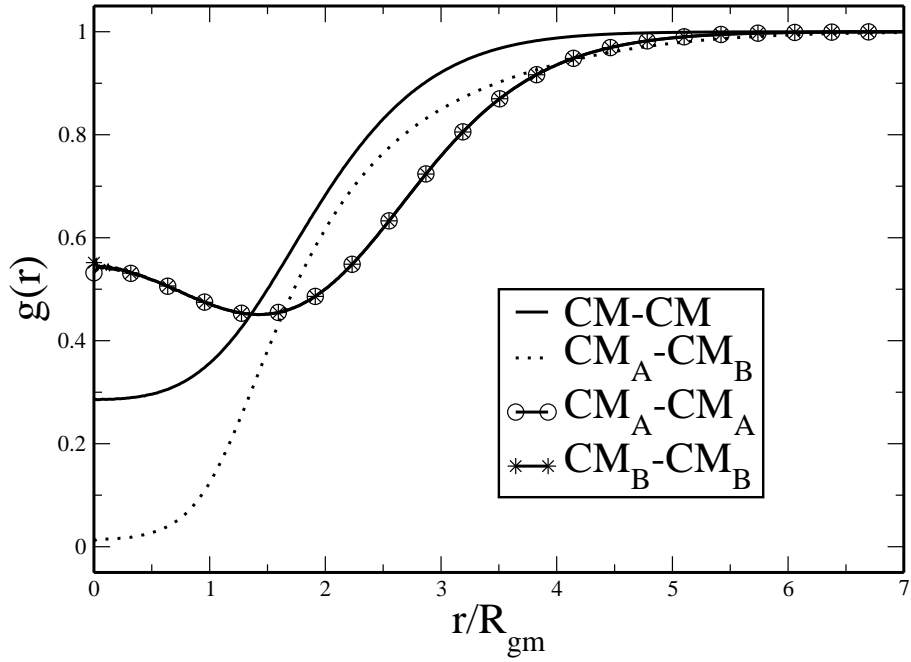}
\caption{Zero density limit of the $CM$ pair distribution functions 
$g_{AA}(r),g_{AB}(r), g_{BB}(r)$ and $g_{CC}(r)$ versus $r/R_g$.  
The results are from MC simulations of $2M$=500 copolymers.}
\label{fig6}
\end{center} 
\end{figure}
 These results together with the previously
 determined intramolecular pair distribution function $s_{AB}(r)$, are
 then inverted to yield effective, monomer-averaged pair potentials
 $v_{AA}(r)=v_{BB}(r)$, $v_{AB}(r)$ and $v_{CC}(r)$.  While the
 inversion of $g_{CC}(r)$ to yield $v_{CC}(r)$ is trivial;
 \begin{eqnarray}
\label{eq7}
v_{CC}(r)=-k_BT \ln g_{CC}(r)
\end{eqnarray}
the inversion of $g_{AA}(r)$ and $g_{AB}(r)$ to yield $v_{AA}(r)$,
$v_{AB}(r)$, given $s_{AB}(r)$, is considerably more complicated,
since one is in effect facing a four-body problem.  For site-site
interaction models of ``diatomics'', the exact relation between
$g_{\alpha\beta}(r)$ and $v_{\alpha\beta}(r)$, for a given $s_{AB}(r)$
is in the low density limit\cite{ladan}:

\begin{multline}
\label{eq8a}
\lim_{\rho\to0} h_{AA}(\mathbf{r})=f_{AA}(\mathbf{r}) +
\left[1+f_{AA}(\mathbf{r})\right]\left\{ \int d\mathbf{x}
\left[f_{AB}(\mathbf{x}) s_{BA}(\mathbf{x-r}) + s_{AB}(\mathbf{x})
f_{BA}(\mathbf{x-r}) \right]\right.\\ + \int d\mathbf{x} \int
d\mathbf{y} s_{AB}(\mathbf{x}) s_{BA}(\mathbf{y-r}) \left[
f_{BB}(\mathbf{x-y}) + f_{AB}(\mathbf{y}) f_{BB}(\mathbf{x-y}) +
f_{BA}(\mathbf{x-r}) f_{BB}(\mathbf{x-y}) \right. \\
\left. \phantom{\int} \left. + f_{AB}(\mathbf{y})
f_{BA}(\mathbf{x-r}) + f_{AB}(\mathbf{y}) f_{BA}(\mathbf{x-r})
f_{BB}(\mathbf{x-y}) \right] \right\}
\end{multline}
\begin{multline}
\label{eq8b}
\lim_{\rho\to0} h_{AB}(\mathbf{r})=f_{AB}(\mathbf{r}) +
\left[1+f_{AB}(\mathbf{r})\right]\left\{ \int d\mathbf{x}
\left[f_{AA}(\mathbf{x}) s_{AB}(\mathbf{x-r}) + s_{AB}(\mathbf{x})
f_{BB}(\mathbf{x-r}) \right]\right.\\ + \int d\mathbf{x} \int
d\mathbf{y} s_{AB}(\mathbf{x}) s_{AB}(\mathbf{y-r}) \left[
f_{BA}(\mathbf{x-y}) + f_{AA}(\mathbf{y}) f_{BA}(\mathbf{x-y}) +
f_{BB}(\mathbf{x-r}) f_{BA}(\mathbf{x-y}) \right. \\
\left. \phantom{\int} \left. + f_{AA}(\mathbf{y})
f_{BB}(\mathbf{x-r}) + f_{AA}(\mathbf{y}) f_{BB}(\mathbf{x-r})
f_{BA}(\mathbf{x-y}) \right] \right\}.
\end{multline}
where $h_{\alpha\beta}(\mathbf{r})=g_{\alpha\beta}(\mathbf{r})-1$ and the $f_{\alpha\beta}(\mathbf{r})$ are the Mayer functions
\begin{equation}
\label{eq9}
f_{\alpha\beta}(\mathbf{r})=f_{\beta\alpha}(\mathbf{r}) = \exp \left[-\beta
  v_{\alpha\beta}(\mathbf{r})\right] - 1,
\end{equation}
The inversion procedure now amounts to solving the coupled integral
equations \ref{eq8a} and \ref{eq8b} for the $f_{\alpha\beta}(r)$,
using the $h_{\alpha\beta}(r)$ and $s_{\alpha\beta}(r)$ from the MC
simulations as input.  Once the $f_{\alpha\beta}(r)$ have been
calculated by an appropriate iterative solution of the two coupled
integral equations, the effective intermolecular potentials
$v_{\alpha\beta}(r)$ follow from equation \ref{eq9}.  The numerical
solution of equations \ref{eq8a} and \ref{eq8b} is facilitated by the
fact that all integrals on the r.h.s. are convolution integrals, and
hence easily evaluated by Fourier transformation, except the
last term of the double integrals, which involves five factors and
corresponds to a fully connected ``bridge'' diagram, which cannot be
resolved by Fourier transformation.
\begin{figure}[!htp]
\begin{center}
\includegraphics[width=12cm,angle=0]{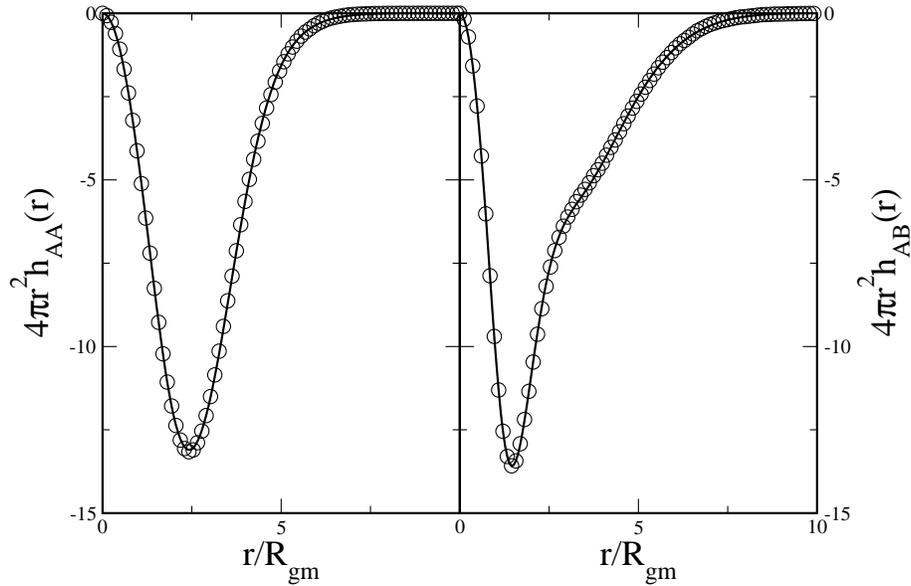}
\caption{ Optimal representation of the MC data for $4\pi r^2h_{AA}(r)$ and $4\pi r^2h_{AB}(r)$ of the ISI model in the zero density limit, achieved by the inversion procedure of equations \protect\ref{eq8a},\protect\ref{eq8b} and \protect\ref{eq9}.  The unknown $f_{\alpha\beta}(r)$ are parametrised by sums of 10 Gaussian functions.} 
\label{fig7}
\end{center} 
\end{figure}
\begin{figure}[!htp]
\begin{center}
\includegraphics[width=12cm,angle=0]{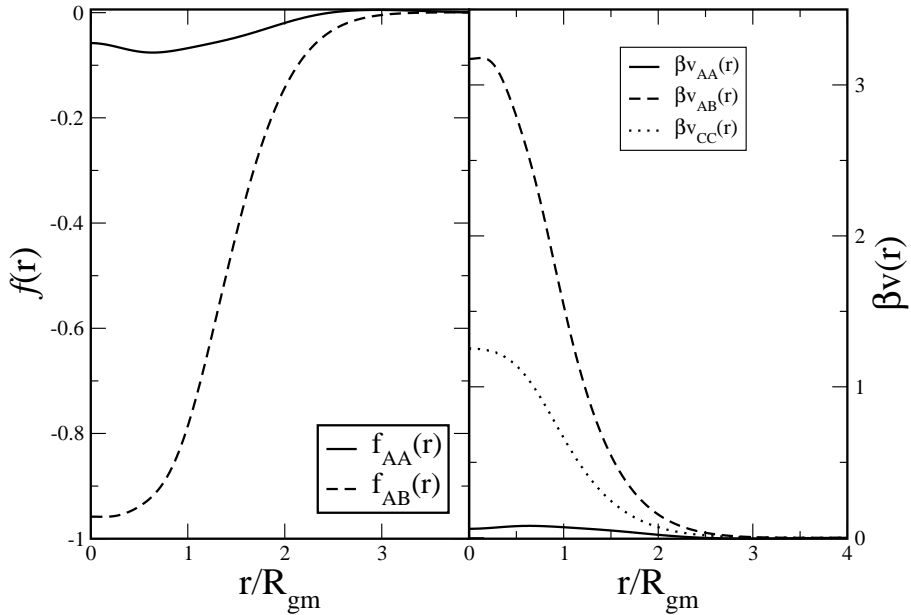}
\caption{{\bf Left:} Mayer functions $f_{AA}(r)=f_{BB}(r)$ and
$f_{AB}(r)$ versus $r/R_{gm}$, as derived by the inversion procedure
outlined in the text and in figure \protect\ref{fig7}. {\bf Right:}
Resulting effective pair potentials $v_{AA}(r)=v_{BB}(r)$, $v_{AB}(r)$ and $v_{CC}(r)$, versus $r/R_{gm}$; the effective potentials correspond to
the zero density limit. }
\label{fig8}
\end{center}
\end{figure}
If the bridge term is left out, in the spirit of the familiar
hyper-netted chain (HNC) approximation\cite{hans2}, major difficulties
are encountered when attempting a numerical solution, because of the
inconsistency introduced by neglecting the bridge term, which is a
violation of the connectivity constraints.  However the task of
including the bridge term may be achieved by noting that all integrals
in equations \ref{eq8a} and \ref{eq8b} may be calculated analytically,
if $s_{AB}(r)$, $f_{AA}(r)$, $f_{AB}(r)$ are sums of Gaussian
functions centred on $r=0$.  We have hence fitted the MC data for
$s_{AB}(r)$, shown in figure \ref{fig2}, by a sum of four Gaussian
functions.  The (two) unknown Mayer functions $f_{AA}(r)=f_{BB}(r)$,
$f_{AB}(r)$ are represented by a sum of 10 Gaussian functions, and the
amplitudes and widths are varied until the resulting pair correlation
functions $4\pi r^2h_{\alpha\beta}(r)$ yield the best fits to the MC data as
illustrated in Fig \ref{fig7}.  The corresponding optimal Mayer
functions $f_{AA}(r)$, $f_{AB}(r)$ and the resulting effective pair
potentials $v_{AA}(r)$ and $v_{AB}(r)$ are plotted in Fig \ref{fig8}.
As expected, $v_{AA}(r)=v_{BB}(r)$ is small compared to $k_BT$ for all
site-site distances, while $v_{AB}(r)$ is roughly Gaussian in shape,
with an overlap ($v_{AB}(r=0)$) value of 3.2$k_BT$.  The effective
pair potential $v_{CC}(r)$ between the overall copolymer $CM$s, calculated from
eq. \ref{eq6} and the MC data for $g_{CC}(r)$, is also shown in Fig
\ref{fig8}.  Its amplitude at full overlap, $v_{CC}(r=0)\approx
1.25k_BT$ is almost three times smaller than $v_{AB}(r=0)$ but its
range is significantly wider, as one may expect, since the effective
potential $v_{CC}(r)$ involves an implicit averaging over relative
orientations of the elongated diblock copolymers.  It is interesting
to note that the entropic barrier of 3.2$k_BT$ for complete overlap of
the mutually avoiding strands A and B of the two copolymers is
substantially higher than found for homopolymers, where $v(r=0)\approx
1.8k_BT$ for sufficiently long
polymers\cite{hall}\cite{bolh}\cite{pelis}. This is a consequence of
the fact that monomers of the same strand can overlap (ideal
polymers), and that the strands are hence more compact than the
swollen self-avoiding walk polymers.  Interestingly, the effective
potential $v_{AB}(r)$ for two copolymers is rather close to that found
for binary mixtures of untethered A and B polymers\cite{add3}.

\section{Conclusion}
We have introduced and investigated a highly simplified model of a
symmetric diblock copolymer which leads to microphase separation, the
ISI model.  The binary mixture counterpart of ISI, where A and B
strands are untethered is the macromolecular equivalent of the
Widom-Rowlinson model\cite{widom2}, and like the latter leads to
macroscopic phase separation\cite{add3}.  Despite the simplicity of
the ISI model, a fully quantitative investigation of its structure,
thermodynamics and phase behaviour would still be a daunting task, and
only partial results have been reported in the present paper.  In
order to go further we propose a coarse-graining strategy, whereby an
AB block copolymer is schematized by two ``blobs'' tethered by an
entropic spring, deriving from an effective intramolecular pair
potential $\phi_{AB}(r)$ between the $CM$s of the two strands.  We were able
to extract $\phi_{AB}(r)$ and the intermolecular potentials $v_{AA}(r)$ and
$v_{AB}(r)$ from the site-site pair distribution functions $s_{AB}(r)$,
$g_{AA}(r)$, $g_{AB}(r)$ in the zero density limit, as generated by MC
simulations for an isolated pair of interacting copolymers with 2M
monomers ($10^2 \lesssim M \lesssim 10^3$).  The simulation of the
resulting coarse-grained ``soft diatomic'' representation of the ISI
model is orders of magnitude faster, since the number of degrees of
freedom is reduced by a factor M. Preliminary investigations show that
the coarse-grained model indeed leads to phase behaviour very similar
to that of the original ISI model.  Note that, by construction, the
``soft diatomic'' model will lead back to the exact pair distribution
functions of the ISI, at least at zero density.  From our
earlier experience with effective pair potentials between the $CM$s of
homopolymers, we expect some concentration dependence of the effective
potentials for AB copolymers.  In the case of the former, the
inversion of the single pair distribution function at finite polymer
concentration is straightforward, via the HNC closure which is
extremely accurate for soft potentials\cite{bolh}\cite{krak}.  A
similar inversion scheme for diblock copolymers could in principle be
based on the ``reference interaction site model'' (RISM) formalism\cite{chand},
with HNC closure, but the inconsistencies of this formalism are well
documented, and will require careful attention for its implementation
in an inversion procedure. Since no dramatic density-dependence of the
effective potentials is expected, at least for the athermal ISI model,
an excellent first approximation will be to use the zero density
potentials derived in the present paper to investigate the behaviour
of dilute and semi-dilute copolymer solutions.  A severe check will be
provided by a direct comparison of the pair distribution functions
obtained at finite densities with the effective ``soft diatomic''
model, and from the full underlying ISI model.  Work along these lines
is in progress.  The extension of the present work to asymmetric
($M_A\neq M_B$) diblock, triblock and more complex copolymer systems
is, at least in principle, straightforward.

\section{acknowledgements}
C. Addison wishes to thank the EPSRC for a studentship, and A. Louis is grateful to thank the Royal Society for a university research fellowship.  The authors found constant inspiration in the ideas and papers of Ben Widom.


\begin{thebibliography}{99}
\bibitem{likos} C.N. Likos, Phys.Rep. {\bf 348}, 267(2001)
\bibitem{louis} A.A. Louis, Philos.Trans.R.Soc.London, ser.A{\bf 359}, 939 (2001)
\bibitem{hans1} J.-P. Hansen and H. L\"{o}wen, in ``Bridging time scales: molecular simulations for the next decade'', edited by P. Nielaba, M. Mareschal and G. Ciccotti (Springer Verlag, Berlin 2002)
\bibitem{hall} J. Dautenhahn and C.K. Hall, Macromolecules{\bf 27}, 5399 (1994)
\bibitem{bolh} P.G. Bolhuis, A.A. Louis, J.-P.  Hansen, and E.J. Meijer, J.Chem.Phys. {\bf 114}, 4296 (2001).
\bibitem{krak} V. Krakoviack, J.-P. Hansen, and A.A. Louis, Phys.Rev.E, {\bf 67}, 041801 (2003)
\bibitem{Guen} G. Yatsenko, E. J. Sambriski, M. A. Nemirovskaya, and M. Guenza, PRL {\bf 93}, 257803 (2004)
\bibitem{jusu} C.N. Likos, H. L\"{o}wen, M. Watzlawek, B. Abbas, O. Jucknischke, J. Allgaier and D. Richter, Phys.Rev.Let  {\bf 80(20)}, 4450 (1998) 
\bibitem{pelis} A. Pelissetto and J.-P. Hansen, J.Chem.Phys., {\bf 122}, 134904, (2005)
\bibitem{hans2} J.-P. Hansen and I.R. McDonald, ``Theory of Simple Liquids'', $2^{nd}$ edition (Academic Press, London 1986)
\bibitem{widom1} J.S. Rowlinson and B. Widom, ``Molecular Theory of Capillarity (Clarendon Press, Oxford 1982)
\bibitem{widom2} B. Widom and J.S. Rowlinson, J.Chem.Phys.{\bf 52}, 1670 (1970)
\bibitem{bates} F.S. Bates and G.H Frederickson, Ann.Rev.Phys.Chem.{\bf 41}, 525 (1990) 
\bibitem{binder} K. Binder, Adv.Polym.Sci,{\bf 112}, 181 (1994)
\bibitem{add1} For a recent discussion, see C.I. Addison, A.A. Louis and J.P. Hansen, J.Chem.Phys.{\bf 121}, 612 (2004)
\bibitem{olaj} O.F. Olaj, B. Neubauer and G. Zifferer, Macromol.Theo.Simul.{\bf 7}, 171 (1998) 
\bibitem{lal} M. Lal, Molec.Phys.{\bf 17}, 57 (1969) 
\bibitem{madras} N. Madras and A.D. Sokal, J.Stat.Phys.{\bf 50}, 109 (1988)
\bibitem{li} see {\it e.g.}\ B. Li, N. Madras and A.D. Sokal, J.Stat.Phys.{\bf 80}, 61 (1995)
\bibitem{add2} C.I. Addison, A.A. Louis and J.-P. Hansen, to be published (2005) 
\bibitem{biben} T. Biben, J.-P. Hansen and J.L. Barrat, J.Chem.Phys.{\bf 98}, 7330 (1993)
\bibitem{ladan} B.M. Ladanyi and D. Chandler, J.Chem.Phys. {\bf 62} 4308 (1975)
\bibitem{add3} C.I. Addison, P.A. Artola, J.-P Hansen and A.A. Louis, to be published (2005)
\bibitem{chand} For a review, see D. Chandler, in ``Studies in Statistical Mechanics'', vol 8, P.275, edited by J.L Lebowitz and E.W. Montroll (North Holland, Amsterdam, 1982)



\end{thebibliography}
\end{document}